# Applications of Artificial Intelligence in Particle Radiotherapy

Chao Wu[1], Dan Nguyen[2], Jan Schuemann[3], Andrea Mairani[4], Yuehu Pu[1], Steve Jiang[2,a)]

[1]Shanghai Advanced Research Institute, Chinese Academy of Sciences, Shanghai, China
[2]Medical Artificial Intelligence and Automation (MAIA) Laboratory, Department of Radiation Oncology, University of Texas Southwestern Medical Center, Dallas, TX, USA
[3]Departmet of Radiation Oncology, Massachusetts General Hospital and Harvard Medical School, Boston, MA, USA
[4]Heidelberg Ion Beam Therapy Center, Heidelberg University Hospital, Heidelberg, Germany.

a) Author to whom correspondence should be addressed. Email: Steve.Jiang@UTSouthwestern.edu

***Abstract*—Radiotherapy, due to its technology-intensive nature and reliance on digital data and human-machine interactions, is particularly suited to benefit from artificial intelligence (AI) to improve the accuracy and efficiency of its clinical workflow. Recently, various artificial intelligence (AI) methods have been successfully developed to exploit the benefit of the inherent physical properties of particle therapy. Many reviews about AI applications in radiotherapy have already been published, but none were specifically dedicated to particle therapy. In this article, we present a comprehensive review of the recent published works on AI applications in particle therapy, which can be classified into particle therapy treatment planning, adaptive particle therapy, range and dose verification and other applications in particle therapy. Although promising results reported in these works demonstrate how AI-based methods can help exploit the intrinsic physic advantages of particle therapy, challenges remained to be address before AI applications in particle therapy enjoy widespread implementation in clinical practice.**

## I. Introductions

IN the past several years, a new rising tide of artificial intelligence (AI) has been changing our world in many aspects, including healthcare. This new dawn of AI is mainly attributable to the availability of big data in the internet era, the easy access to graphics processing unit (GPU) based high-performance computing, and the emergence of deep learning (DL) algorithms [1]. Given its broad and exponentially growing applications in medicine, AI has the potential to fundamentally transform the way medicine is practiced [1,2].

As the results of recent technological advances in computer processing power and imaging quality, radiotherapy has become increasingly precise and efficient at the cost of the growing complexity, which has resulted in a near-complete dependence on human-machine interactions [2].The development of highly sophisticated radiotherapy techniques includes intensity-modulated radiation therapy (IMRT), volumetric modulated arc therapy (VMAT), image guided radiation therapy (IGRT), stereotactic body radiation therapy (SBRT)), as well as particle therapy. In radiotherapy, the typical patient treatment workflow includes multiple complex tasks like patient image acquisition, segmentation of tumor target volumes and organs at risk (OARs), treatment planning, patient positioning and immobilization, treatment delivery, and post-treatment follow-up [2,3]. Despite technological advancements, many of these tasks still require time-consuming digital data processing and manual input by radiation oncology professionals [2]. The technology-intensive nature of radiotherapy and its reliance on digital data and human–machine interactions make this discipline particularly suited to benefit from AI to improve the accuracy and efficiency of its clinical workflow [4].

The main rationale of particle therapy is the physical characteristics of the depth dose curve, which has a dose peak (Bragg peak) followed by a sharp fall-off at a well-defined depth in tissues [5,6]. This physical advantage allows particle therapy to achieve higher dose conformity for the tumor volume with a lower dose to the surrounding healthy tissues than photon radiotherapy [6,7]. The integral dose for proton therapy can be a factor three lower than for standard photon treatments [6]. Nonetheless, this physical advantage also imposes a stronger requirement on the accuracy in predicting the range of the treatment field, which makes precise coverage of the target volume in particle therapy more challenging than in photon therapy [8]. Therefore, compared with photon radiotherapy, higher precision and smaller uncertainties are needed for particle therapy, which may be achieved by applying AI techniques to particle therapy. Aspects that can benefit from AI include converting dual energy CT images or CBCT images to stopping power ratio maps, converting MR images to CT images for accurate dose calculation, converting CBCT images to CT images for adaptive re-planning, performing accurate and (near-) real time dose calculations, and verifying range and dose using positron activity distributions, prompt gamma images, secondary electron bremsstrahlung X-ray images, acoustic signals, and luminescence images in water.

Recently, various AI methods have been successfully developed to exploit the benefit of the inherent physical properties of particle therapy. Many reviews about AI applications in radiotherapy have already been published [1,2,3,4,8,9,10], but none were specifically dedicated to particle therapy. We therefore conducted a literature review of recent important papers and synthesize these papers into a comprehensive review of the current status of AI applications in particle therapy. A summary of these AI applications in particle therapy is presented in Table I. The table shows that AI in particle radiotherapy is a new field as some topics have only recently been studied with AI and thus feature a limited number (as low as 1) published manuscripts. However, we believe those topics deserve to be included in this review. AI research in radiotherapy is a rapidly growing field and more and more researchers will join this field. This article, even with limited literature on certain topics, will offer researchers a clear picture on the current status of using AI to improve particle therapy and perhaps more importantly, point to those specific topics where there is not yet much work done and, thus, still room for future research.

In Sections 2 to 5, we present a brief review of different major research areas where AI methods are applied to particle therapy. We have chosen to classify the reviewed papers into four different categories: particle therapy treatment planning, adaptive particle therapy, range and dose verification and other applications in particle therapy. In Section 6, we summarize the review paper and provide more in-depth discussions on the challenges and future of AI in particle therapy.

TABLE I.
Summary of the reviewed AI based applications in particle therapy in this work.

| Application | Tumor Site | AI Method | Reference |
|---|---|---|---|
| SPR mapping from DECT | Head and Neck | Residual Attention GAN | 15 |
| SPR mapping from DECT | Head and Neck | Cycle-GAN | 11 |
| MR-based planning | Liver | Cycle-GAN | 23 |
| MR-based planning | Prostate | Cycle-GAN | 24 |
| MR-based planning | Brain | GAN | 44 |
| MR-based planning | Head | U-Net variant | 45 |
| MR-based planning | Head | U-NET | 46 |
| MR-based planning | Skull base | Cycle-GAN | 47 |
| MR-based planning | Brain | Conditional GAN | 48 |
| Dose calculation | Head and Neck | 3D-CNN | 54 |
| Dose calculation | Lung | LSTM | 55 |
| Dose calculation | Head and Neck\Prostate\Lung\Liver | U-Net variant | 7 |
| CBCT to CT conversion | Prostate | Cycle-GAN | 61 |
| CBCT to CT conversion | Pelvis | U-Net variant | 62 |
| CBCT to CT conversion | Prostate | U-Net | 63 |
| CBCT to CT conversion | Head and Neck | U-Net variant | 57 |
| CBCT to CT conversion | Head and Neck | U-Net variant | 64 |
| CBCT for SPR mapping | Head and Neck | Cycle-GAN | 65 |
| Range verification (PET) | \ | LSTM | 75 |
| Range verification (PET) | \ | LSTM, BiLSTM, GRU, BiGRU and Seq2seq | 76 |
| Range verification (PET) | \ | RNN | 77 |
| Range verification (PET) | \ | GAN | 78 |
| Range verification (PG imaging) | \ | U-Net | 95 |
| Range verification (SEB X-ray) | \ | U-Net | 102 |
| Range verification (Acoustic signals) | \ | Bi-LSTM | 108 |
| Range verification (Luminescence water image) | \ | U-Net | 109 |
| MU prediction | \ | Random forest\ XGBoost \Cubist | 112 |
| MU prediction | \ | GPR\SNN | 113 |
| Delivery error prediction | Prostate | Linear regression\Random forest\Neural network | 116 |
| Treatment modality selection | Head and Neck | ML classifiers | 117 |

## II. AI IN PARTICLE THERAPY TREATMENT PLANNING

### A. *Stopping power ratio mapping from dual energy CT*

Dose calculation based on computed tomography (CT) images in particle therapy requires appropriate conversion from the Hounsfield Units (HU) in CT images to the corresponding stopping power ratio (SPR) [11,12]. One common way to obtain SPR is using a piecewise linear fit between SPR and the corresponding HU number of single energy CT (SECT) based on measurements for tissue substitutes with known properties [11,13,14]. However, using this empirical fit directly for HU-RSP calibration may cause inaccuracy since tissues may have the same CT number but different SPR or vice versa. [11,15]. Moreover, due to the approximation of real tissue with tissue substitutes used for measurements, the patient-specific tissue variations may also lead to errors in SPR calibration [11,14]. Because of its material differentiation ability, dual energy CT (DECT) has recently been utilized to estimate the SPR. Methods that derive SPR from DECT images have been proposed by several research groups [16-20]. Among these methods, the physics-based method that uses physical equation to derive the SPR maps from DECT images in a voxel-wise manner is mainly used in clinical routine [18-20].

Despite its advantages, DECT has not been widely utilized in hospitals like the standard single energy CT (SECT), mainly owing to its implementation and operation costs [15]. Moreover, DECT suffers from extra noise and artifacts caused by the sub-optimal implementations of DECT acquisition [11]. Considering these limitations in current DECT implementation, Charyyev et al. proposed an alternative approach adopting AI based methods to generate synthetic DECT (sDECT) from SECT and derive SPR from sDECT for dose calculation [15]. They developed a residual attention generative adversarial network (GAN) [21] to synthetize DECT images from SECT for proton therapy. This method can generate synthetic low energy CT (sLECT) and synthetic high energy CT (sHECT) from SECT images. The corresponding SPR maps generated from the sDECT are quantitatively close to those from original DECTs in general, with reduced noise level and artifacts.

When adopting the physics-based mapping method that has already been used in clinical routine [15], the DECT-derived SPR maps could potentially be compromised by noise and artifacts in the DECT images. This could lead to uncertainties and inaccuracies in the subsequent clinical applications [11]. To address this issue, Wang et al. integrated a residual attention architecture into cycle consistent generative adversarial networks (cycle-GANs) [22] to predict SPR maps from DECTs for proton therapy. The results showed the proposed model can accurately predict SPR maps from DECT images.

The datasets used in the abovementioned work of Charyyev et al. [15] and Wang et al. [11] both include 20 head-and-neck patients. Leave-one-out cross validation that repeatedly leaves one test patient out and trains the model by the remaining 19 patients was conducted in both studies. Since the patient datasets feature high anatomical complexity and variability between patients, a comprehensive evaluation with a larger number of patients and diverse tumor sites should be considered in future studies to further reduce model bias. Moreover, datasets from different institutions could also be considered to evaluate the generalization capability these two methods.

### B. *MR-based planning*

Radiation therapy relies on CT as the image modality for daily patient setup, treatment planning and dose calculation [8,23,24]. However, delineation of clinical target volumes (CTVs) based on CT images continues to be a weak point of CT due to its poor soft tissue contrast [23,24]. This issue and resulting uncertainties may have a bigger impact on the outcome of particle therapy than that of photon therapy [24]. Magnetic resonance imaging (MRI) is often proposed as a complementary imaging modality to CT because of its excellent high soft-tissue contrast [23]. However, the separate acquisition of CT and MR images leads to a new level of inherent complexity and uncertainty in the MR-CT pair registration process [23,24]. MR-only treatment planning has been suggested as a solution because it can achieve accurate delineation and spare the patient from CT radiation exposure, while avoiding the inherent registration errors between CT and MR images in CT based treatment planning [24]. One major task in MRI-only treatment planning is to generate synthetic CT (sCT) images that can be used as CT substitute for dose calculation because MRI signals are not explicitly linked to CT HU numbers which can be used to generate SPR for dose calculation [23,24].

The available sCT generation methods that can be broadly categorized into atlas-based [25-31], segmentation-based [32-36] and machine learning-based methods [37-43]. Atlas-based methods often fail to handle atypical patient anatomy and suffer from registration errors [24,44]. Segmentation-based methods are limited by the time-consuming multiple MRI sequences acquisition and manually contouring process [24]. Machine learning-based methods mainly include dictionary learning-based methods [37], random forest [39] and deep learning methods [40-43]. Unlike the other two methods, deep leaning methods can eliminate the need for handcrafted features by allowing the deep network to learn its own optimal features [23]. Spadea et al. first adopted a convolutional neural network (CNN) for the sCT generation in MRI only proton therapy. 15 pairs of MRI/CT head scans were used to train U-Nets to predict the CT HU from MRI intensities for each voxel [45]. Neppl et al. compared the performance of utilizing two-dimensional (2D) and three-dimensional (3D) U-Net for sCT generation from head MRIs [46]. Mean absolute and mean error (MAE/ME) results revealed a slight advantage of 2D U-Net over 3D U-Net, while similar high dose calculation accuracy was achieved by both 2D and 3D U-Net.

One limitation of the CNN-based methods is that the sCTs accuracy may be deteriorated by misalignment between CT/MRI pairs. As an alternative network, GANs have also been introduced for sCT generation. Liu et al. used a 3D dense cycle-GAN to generate abdominal and pelvic sCTs for MRI-based proton treatment planning [23,24]. In both studies, the authors integrated the dense block into a 3D cycle-GAN framework to capture 3D spatial information and effectively learn the nonlinear mapping between MRI and CT pairs. In parallel, Erfani et al. also adopted a 3D cycle-GAN for sCT generation for skull-based tumors [47]. Samaneh et al. demonstrated the feasibility of using GANs for sCT generation for brain IMPT with robust optimization [44].

Conversion of MR to sCT is of importance in radiotherapy in general. However, particle therapy dose calculation requires more accurate CT numbers in sCTs than photon-based radiotherapy. Therefore, for particle therapy, dosimetric evaluation of the generated sCTs is more important and should be carefully conducted. For a heterogeneous set of imaging protocols, Maspero et al. accessed the feasibility of dose calculation based on sCTs that were generated by a conditional GAN, which was trained with 60 pediatric brain patients with a large range of size, shape, and age [48]. Analysis of image similarity and dosimetric results showed that accurate MR-based dose calculation was achieved using the proposed method, even when using a heterogeneous patient dataset.

One limitation of these studies is the relatively small dataset used to train the models, future work should consider using large and heterogeneous datasets to fully evaluate the performance of various sCT generation approaches. Furthermore, since the sCTs generated from these approaches need to be further converted to SPR maps for dose calculation, future work can include approaches which extends the current methods to realize a direct mapping from MRI to SPR for dose calculation purpose.

### *C. Dose calculation using deep learning*

Recent studies have demonstrated the feasibility of using convolutional neural network (CNN)-based dose calculation methods for IMRT or VMAT [49-53]. For particle therapy, Nomura et al. proposed a spot-scanning proton dose calculation method using a three-dimensional convolutional neural network (3D-CNN) [54]. The dataset used for model training and evaluation includes 193 head and neck squamous cell carcinoma patients. The proposed 3D-CNN model can calculate volumetric proton dose distribution of each single beam spot from variable spot data (initial beam energy, spot weight and spot position), SPR volume and the binary surface mask which indicates whether a voxel is inside the irradiated object or not. The total dose calculation time is around 0.8 seconds for a plan using 1,500 spots with a consumer grade GPU. The authors also used stochastic batch normalization (SBN) to calculate the model's uncertainty as a confidence measure of dose calculation. Furthermore, the trained 3D-CNN model can be fine-tuned efficiently for proton dose distributions calculated with other beam parameters or calculation methods using a small amount of training data by using the transfer learning technique. One limitation of this work is that the voxel resolution of the dose data used for training and evaluating was set to 4 mm due to the limited GPU memory capacity, while for clinical use, the voxel resolution should be at least 2 mm. Neishabouri et al. proposed a further approach to calculate the dose, where they designed a model based on Long-Short Term Memory (LSTM) networks to calculate the dose for each single pencil beam [55]. The proposed LSTM model was trained/tested on two separate data sets. The authors first successfully trained their model on phantom data set with sizable artificial high SPR heterogeneities. Subsequently, they trained/tested their model on lung patient cases exhibiting highly pronounced inhomogeneities, e.g., low SPR lung tissues and high-density rib cage. Their model was able to successfully predict the bimodal Bragg peak behind heterogeneous interfaces. Calculating the dose in a single pencil beam level allows the generation of the dose influence matrix required for dose optimization routines and therefore, can easily be integrated into today's treatment planning systems. The reported dose calculation run time goes as low as 1.5 ms for a single pencil beam and a consumer-grade GPU. However, while they illustrate the generalization of their model to other proton energies, they have not shown yet the performance of the model for a full proton plan.

CNNs have also been used to improve the accuracy of proton dose calculation by predicting Monte Carlo (MC) dose distributions from analytical pencil beam (PB) dose distributions for different tumor sites [7]. The proposed model is based on the hierarchically densely connected U-Net network [52,56]. This model uses the PB dose and patient CT images as inputs to generate a dose distribution equivalent to MC simulations. The patient dataset includes 290 patients (90 head and neck, 93 liver, 75 prostate, and 32 lung cancer) that were treated with double scattered proton beams. For each tumor site, four numerical experiments were implemented to explore various combinations of training datasets. The results showed that training the model on data from all tumor sites together and using the dose distribution of each individual beam as input yielded the best performance for all the four tumor sites. The predicted dose distributions show significant improvement over the PB dose distributions in all the considered evaluation criteria (Gamma Index, mean square error (MSE), dose volume histogram (DVH) and dose difference histogram). Moreover, the trained model can be deployed to different hospitals through transfer learning, even when using a different treatment plan delivery, i.e., IMPT vs. double scattering. The average dose prediction time of the proposed model for one single field is less than 4 seconds. The authors claimed that they did not dedicate any efforts to improve the model efficiency, which can be achieved through methods like model compression. To date, the fastest proton PB dose calculations are in the sub-second range. The total time required to obtain the converted MC dose distribution can potentially be within a second. Therefore, such deep learning based methods have the potential to achieve (near-) real time dose calculation efficiency and may have a role in certain clinical applications. In addition, the deep learning-based methods are fundamentally different from MC methods and can be used as secondary dose calculation tools for quality assurance purposes. Once the accuracy and efficiency of these methods have been thoroughly demonstrated, users may implement DL models as an add-on to their existing PB based planning system, without acquiring and commissioning an MC based dose engine.

## III. AI FOR CBCT CONVERSION IN ADAPTIVE PARTICLE THERAPY

Adaptive radiotherapy (ART) intends to achieve the best possible target coverage and OAR sparing in the presence of patient anatomy changes [57,58]. To that end, ART utilizes an imaging feedback loop to monitor and quantify patient specific anatomic and/or biological changes, access the actual delivered dose and modify the treatment plan accordingly [57,58]. For particle therapy, the necessity for ART is greater than for photon therapy as the steep dose gradients make particle therapy more sensitive to anatomical changes during the course of fractionated therapy [59,60].

Imaging feedback loop during the treatment course is an essential part in ART. Because the actual delivered dose that calculated from these images can be used to access the necessity for adjusting the treatment plan [57,58]. Recently, on-board cone-beam computed tomography (CBCT) systems have become available in many particle therapy centers. CBCT is used for accurate pre-treatment patient alignment and to obtain information about inter-fractional patient changes [57, 58]. However, the image quality of CBCT is typically not sufficient for dose calculation because the inaccuracy of the reconstructed HU numbers in CBCT images due to various artifacts [59]. Therefore, raw CBCT images cannot be directly used for clinical dose calculations and are currently solely used for accurate patient alignment. Similar to the use of AI methods that convert MRI images into sCT images for dose calculation, various AI methods have been adopted to make the CBCT images suitable for dose calculation and plan adaptation in particle therapy [57,59,61,62,63]. The AI related research work for particle therapy can be divided into two main categories: CBCT to CT conversion [57,59,61,62,63,64] and CBCT to SPR mapping [65]. These works will be reviewed in the following two sub sections.

## A. *CBCT to CT conversion*

Various CBCT to CT conversion approaches for dose calculations have been published in literature. These approaches can be categorized into simple look-up table (LUT) based approaches [66], histogram matching method [67], deformable image registration (DIR) method [68,69], and scatter correction method [70,71]. Recently, deep learning-based methods have also been developed for CBCT to CT conversion [57,59,61, 62,63,64,71,72,73,74]. Compared with other methods, deep learning-based methods offer the advantages of not relying on a planning CT for sCT generation and rapid image correction [57,59]. Deep learning-based methods include DCNN (supervised training with paired images) and GAN (unsupervised training with unpaired images) [72].

Acquiring aligned paired CT and CBCT images is often difficult in practice. To overcome the limitation of lacking paired CBCT and CT dataset in reality, Liang et al. first proposed to adopt cycle-GAN to synthesize CT images from CBCT images for head-and-neck patients [72]. For particle therapy, Kurz et al. evaluated the performance of using a cycle-GAN to convert CBCT images into sCTs for VMAT and proton therapy [61]. The results showed that the accuracy of dose calculation using the sCTs was sufficient for VMAT, but not for proton therapy.

DCNN that used supervised training with paired images has also been adopted for sCT generation from CBCT in particle therapy. Hansen et al. used pairs of measured and corrected CBCT projections to train a 2D U-Net like model to correct CBCT images for dose calculation and evaluated the accuracy of dose calculations based on the corrected CBCTs for IMPT and VMAT plans [62]. For pelvic patients, the results showed satisfactory dose calculation accuracy for VMAT plans and insufficient accuracy for IMPT plans. Lalonde et al. also trained a U-Net to reproduce MC projection-based scatter correction from raw CBCT projections for head and neck proton therapy [59]. Landry et al. trained a single U-Net using three types of training sets: (1) raw and corrected CBCT projections; (2) raw CBCTs and DIR-synthetic CTs; (3) raw and reference corrected CBCT [63]. For prostate patients, training the model using corrected CBCT as reference achieved the best performance for single-field uniform dose (SFUD) proton plans [63]. Lalonde et al. evaluated the performance of training a U-Net to reproduce MC-based CBCT scatter correction for head and neck adaptive proton therapy [59]. Dosimetric analysis and proton range analysis results showed good agreement with MC simulated scatter free CBCT. Thummerer et al. compared three different methods (DCNN, DIR and analytical image-based correction) to correct CBCTs and create sCTs that are suitable for head and neck proton dose calculations [57]. The results showed that the synthetic CTs created by the DCNN method have the highest image quality. High proton dose calculation accuracy was achieved by both DCNN and DIR based sCTs. Meanwhile, the analytical image-based correction resulted in the lowest image quality and dose calculation accuracy among these three methods. In another comparison study for head and neck patients, Thummerer et al. used a dataset of 27 head and neck patients, containing planning CT, repeat CTs, CBCTs and MRs to train a U-Net variant for sCT generation and compared sCTs performance in terms of image quality and proton dose calculation accuracy [64]. The results showed that CBCT-based sCTs had a higher image similarity when compared to planning CTs images than MR-based sCTs. In terms of dosimetric evaluation results, CBCT-based sCTs and MR-based sCTs seemed to be equally suited for daily adaptive proton therapy.

## B. *CBCT for SPR mapping*

Harms et al. recently proposed to extend cycle-GAN from CBCT to CT conversion to direct CBCT to SPR mapping [65]. With their proposed framework, the SPR map, which is directly used for dose calculation, can be learned from a CBCT image. By deriving the SPR map from a daily CBCT image, the resulting dose distribution based on a patient's current anatomy can be determined. The proposed DL SPR map generation method performs similarly to the DIR-based method for evaluation of day-to-day patient anatomy changes on CBCT for head-and-neck patients. The limitation of this work lies in the HU-SPR curve used to generate the training data, which had an underlying root mean squared error (RMSE) of 5.5% upon validation on various materials [65].

In summary, AI based methods that can learn the nonlinear mapping relationship between CBCT and CT images or CBCT images and SPR maps. These methods have been introduced to adaptive particle therapy for different tumor sites. Promising results have been reported in some of the abovementioned work for anatomical sites like prostate and head and neck. Future work should investigate the generalization of these AI based approaches for more anatomical sites. In some of the abovementioned studies, insufficient dose calculation accuracy was reported for treatment planning based on sCTs [61,62]. Further improvements in dose calculation accuracy might be achieved when using a larger and more inhomogeneous dataset.

## IV. Range and dose verification

The characteristic shape of the Bragg peak in particle therapy enables the use of steep dose gradients at the distal end of the treatment fields, which offers opportunities for a very precise dose delivery to the target with superior sparing of OAR. However, this high conformality also increases the dependency on accurate calculation of the range position or dose distribution [60]. The range uncertainty may originate from several factors such as patient setup errors, organ motion, approximations in dose calculation, and patient anatomical changes [5]. Using outdated images of the patient anatomy for treatment planning is one of the major sources of range uncertainties [60]. Ideally, the treatment plan should be adapted online as soon as anatomical changes occur [60]. The rational for AI -based online verification in particle therapy is that the spatial distribution of particle-induced secondary signals correlates with both dose distribution and the particle range inside the patient [75-78], and this correlation can be precisely learned with AI based techniques. Various online verification methods for particle therapy have been proposed, including positron emitters, prompt gamma (PG), acoustic signal, Cherenkov photons, and luminescence images of water [75,76,78]. AI approaches have been proposed to improve the predictive abilities for all the abovementioned range verification methods.

### *A. Positron activity distribution*

Positron emission tomography (PET) has been used clinically in particle therapy for non-invasive, in-vivo monitoring of the charged particle beam range in patients [79]. The distribution of positron emitters (e.g., $^{15}O$ and $^{11}C$) produced during the irradiation of charged particle beams can be acquired by PET scanners, which correlates with dose distribution [76-79]. Typically, the range or dose verification is obtained by comparing the measured PET distribution with a reference distribution, which can be obtained either from previous measurements [80] or MC simulation [81-84] or analytical modeling [85-88]. On the other hand, methods intend to map measured PET profile to the expected dose profile has also been published [89-90]. Despite some promising results, these methods suffered from extensive computing workload or poor PET counting statistics and image artifacts [89-90].

Recently, AI based approaches have been proposed to learn the nonlinear correlation between the distributions of positron emitters and dose. Li et al. used LSTM recurrent neural network (RNN) regressing model to predict the one-dimensional (1D) dose distribution for mono-energetic beams of different energies and irradiation positions from PET images [75]. Liu et al. compared the performance of using five types of RNN models LSTM, (bidirectional long short-Term memory) Bi-LSTM, (gated recurrent unit) GRU, (bidirectional GRU) Bi-GRU, and Seq2seq to identify the relationship and investigated the impact of including anatomical information (HU numbers in CT images) on the model performance [76]. The results showed that Bi-GRU RNN had the most stable and accurate prediction with good generalization capability, especially with the patient anatomical information. Hu et al. [77] and Ma et al. [78] used two types of models (RNN and GAN) to further extended the two aforementioned work mainly by three aspects: 1) adding the SPR map and patient CT in addition to positron activity distribution to improve model generalizability, 2) realizing verification of range prediction for the center and off-center positions along the fields in three dimensions, and 3) accessing the feasibility of this AI based approach for a spread-out Bragg peak (SOBP) case in addition to mono-energetic beam cases. One limitation of these studies is that the dataset adopted for model training and evaluation were simulated based on a CT image phantom using similar mono-energetic beam energy setting. The proposed models' generalizability for different tumor sites and different beam energy settings needs to be rigorously checked before clinical application. Furthermore, the PET images of different SOBP plateau widths appear to be quite similar in the flat region. This similarity may adversely weaken the underlying correlation between the distribution of positron activity and dose for SOBP cases [77,78].

### *B. Prompt gamma imaging*

During charged particle irradiation, besides positron-emitting isotopes, when a target nucleus is first excited during a scattering event with a charged particle, characteristic photons (PGs) may be emitted [91-94]. Unlike PET, whose data acquisition process is typically performed after the particle beam irradiation due to the relatively long (several minutes) lifetime of the $\beta$+-decay, the PG emission occurs almost instantaneously requiring online detection [95]. Therefore, PG imaging can potentially achieve real-time range and dose verification [94,95]. Although PG images are highly correlated to the dose distribution, the gamma-ray emission distribution is energy dependent and thus different from dose distribution. This means that the particle-induced gamma-ray images cannot directly be used for dose verification [95]. Recently, Schumann et al. proposed to combine a filtering procedure based on Gaussian-power law convolution with an evolutionary algorithm to estimating the dose distribution from PG emission profiles [96]. The feasibility of this filtering approach was demonstrated for a SOBP in a homogeneous water target. However, this approach has not yet been validated for heterogeneous media and patient data.

AI based method recently has been applied to predict dose distribution from PG images for particle therapy [95]. Liu et al. investigated the feasibility of applying a U-net model to fulfill this task [95]. They conducted MC simulations using a 100-MeV proton pencil beam to irradiate 20 digital brain phantoms with different entrance location and beam size. The MC simulated PG and proton dose images were then used for model training and evaluation. The results showed that the predicted pseudo proton dose distributions are in excellent agreement with the simulated ground truths. The limitation of this work is that the accurate dose prediction requires high quality PG images, while in practice, the quality of PG images could be degraded by factors like noise, low counts, and limited spatial resolution.

### *C. Secondary electron bremsstrahlung X-ray image*

Imaging of the secondary electron bremsstrahlung (SEB) X-ray emitted during particle ion irradiation has also been utilized for beam range and width verification [94,97]. SEB X-rays are thought to be more efficient due to a larger emission intensity [94,97]. Low-energy X-ray cameras have been developed and used to image SEB X-ray of proton [98,99] and carbon ion beams [100]. After the improvement of camera performance, real-time imaging of the carbon ion beam has been realized [101]. However, SEB X-ray images are not linearly correlated to dose images. Image conversion methods need to be applied to convert SEB X-ray images to dose images for range and width verification. Moreover, when using X-ray images for range verification, the limited spatial resolution and low-counts of the X-ray camera may introduce errors in the range estimation [102]. To address these issues, AI method has been applied to convert SEB X-ray images to dose images. Yamaguchi et al. proposed to use two U-net models to improve the accuracy of the range and width estimation of SEB X-ray images, the first for SEB X-ray to dose image conversion and the second for improving the spatial resolution of the dose images [102]. Instead of using MC simulations for data generation, a dedicated model function with parameters extracted from the measured data by the X-ray camera was developed to accelerate the whole data generation process. The proposed U-net models trained with these realistic simulated data can accurately predict dose images from both simulated and measured testing data of noisy SEB X-ray images. The predicted dose images can then be used to estimate the ranges and widths of carbon ion beams. Further work should focus on validating the performance for a wider range of beam setup, heterogeneous tissues and including SOBP datasets.

### *D. Acoustic signals*

Methods that exploit thermoacoustic emissions induced by the localized heating of the energy deposition in tissue also have the potential to achieve in vivo range verification [79,93]. Due to the characteristic particle dose deposition, two types of macroscopic pressure waves are emitted by either the pre-peak Bragg curve dose deposition or the Bragg peak dose deposition [103]. Once the arrival time of the acoustic wave from the Bragg peak dose deposition is measured, the time-of-flight (TOF) method can be adopted to calculate the distance between the Bragg peak and the detector [103]. Several recent works have examined the feasibility of using the TOF method of the acoustic signal for range and dose verification in both uniform water medium and heterogeneous tissues [103-106]. Recently, a feasibility dose verification study of applying a time reversal-based reconstruction method in both 2D and 3D heterogeneous tissues has been proposed to address two limitations of the TOF method:1). the complicated extraction of arrival time in the TOF method may potentially lead to poor arrival time accuracy, 2). TOF method cannot provide dose profile information [105]. However, compared with the TOF method, the time reversal method is time consuming and can take up to several minutes for the calculation of a 3D phantom even after GPU acceleration. To address this issue, Yao et al. used a Bi-LSTM to predict 1D dose distributions from proton acoustic waveforms [108]. The proposed framework can identify the correlation between acoustic waveforms and dose profiles in heterogeneous tissues and predict both the Bragg peaks location and dose distributions, while being more computationally efficient. Although this feasibility study showed promising results, various challenges need to be solved before applying this approach to clinical practice, including the potential coupling between the patient and sensor, adjusting to different delivery modes of particle therapy, and how to accurately acquire acoustic parameters for different tissues [108].

### E. Luminescence image of water

The luminescence emitted from water during charged particle irradiation could also be used for range and dose estimation [109-111]. During the irradiation, the luminescence image of water can be obtained by a cooled charge-coupled device (CCD) camera [110,111]. The luminescence of water is thought to be emitted by the same mechanism as Cerenkov light, but generated at lower energy than the Cerenkov light threshold [109-111]. Therefore, a previous study proposed to correct the depth profiles of the measured luminescence images of water by subtracting the simulated distributions of the Cerenkov light for proton therapy [110,111]. However, the Cerenkov light distribution used in this correction method needs to be calculated using MC simulations, which is time consuming. Yabe et al. proposed using AI based method to predict 2D dose distributions from the measured luminescence images of water for protons and carbon ions using a U-Net model [109]. 2D dose distributions and luminescence images of water of protons and carbon ions were utilized to train the model. The results showed that the trained model can efficiently and accurately predict dose distributions from measured luminescence images. The major limitation of this work is that the luminescence image data used in this study were simulated or measured during the irradiation of mono-energetic proton or carbon-ion beams. The performance of the trained model is only validated for some incident energies of protons and carbon ions. Therefore, any deviation from measurement conditions applied in this work require preparation of new training data and model retraining [109]. In addition, the luminescence signal can only be obtained in pure water or from shallow depths in irradiated tissue.

## V. OTHER AI WORK FOR PARTICLE THERAPY

### A. Patient specific QA

Converting the machine deliverable monitor units to the prescribed dose distribution is a vital process for precise treatment delivery in radiation therapy [112,113]. Currently, no commercial TPS is available to calculate monitor units and output factor (dose per monitor units) for passively double-scattered and uniform scanning proton therapy because of the complexity of the beam delivery systems [112, 113]. In the current clinical routine, the field-specific monitor units and output factors are calculated using results of measurements, empirical models (EMs), or MC simulation [112-115]. Recently, AI techniques have been adopted as a secondary check tool for the conventional time-consuming monitor unit or output factor calculation. Bao used data of 1754 treatment fields that were measured in a water phantom in order to train and evaluate the output factor prediction model for a double scattered proton machine (Mevion) [112]. They trained three types of machine learning models (Random forest, XGBoost, and Cubist) and compared the model performance with the clinically used semi-empirical model. Among these three methods, the Cubist algorithm provided the most accurate and robust prediction. Compared with the semi-empirical prediction model, all machine learning methods predicted the output factors for double scatter proton machines with greater accuracy. For uniform scanning proton therapy, Grewal et al. used two MATLAB-based machine and deep learning methods, Gaussian process regression (GPR) and shallow neural network (SNN), for output factor prediction [113]. Four patient quality assurance (QA) parameters (range, modulation, field size, and measured output factor) of 4,231 patient-specific field QA measurements were used for model training and evaluation. The results showed that the GPR and SNN models outperformed the EMs in terms of prediction accuracy.

Besides monitor units and output factor prediction, AI methods have also been adopted for treatment delivery error prediction in proton pencil beam scanning delivery [116]. Three types of machine learning models (linear regression, random forest, and neural network) were trained to predict delivered spot parameters based on planned parameters. The patient dataset used in this study contained planned and delivered spot parameters in treatment plans of 20 prostate patients. The planned spot parameters (spot position, monitor units and energy) were extracted from TPS, while the delivery spot parameters were extracted from irradiation log-files. Dosimetric evaluation results showed that the random-forest-predicted dose distribution was in closer agreement to the delivered dose distribution than the planned dose distribution. One limitation of this study is that the predicted dose distribution was tested on only a single prostate patient, which makes the evaluation process not very convincing. Future work should expand the dataset to fully evaluate the performance of the trained model.

### B. Treatment modality selection

Clinical decision systems (CDS) have the ability to connect a current patient to past decisions made by clinical experts and help inform current assessments [117]. AI based techniques have been adopted to construct a CDS so that clinicians can efficiently find previously approved treatments that are achievable for a new patient. The goal is to aid in the selection of treatment modality and potential dosimetric tradeoffs. Valdes et al. proposed a machine learning approach which can identify previously approved treatment plans which are achievable for a prospective patient to provide patient-specific decisions [117]. In this work, a library of historical photon and proton treatment plans and patient-specific feature sets were used to construct the machine learning classifiers. The main limitation of this work is the limited dataset, this may limit the chance to find an optimal match and reduce the number of dose trade-offs that can be explored.

## VI. Discussion and conclusions

This review presents an overview of AI applications that have been used for particle therapy, synthesizing a total of 30 recent research publications. The studies show that AI techniques have been increasingly applied to achieve full exploitation of the physical advantages of particle therapy over the past few years. As discussed before, the uncertainties related to the finite ion beam range can largely impact the physical advantages offered by particle therapy [93,94]. The sharp dose gradient of the Bragg peaks, by nature, means that any small uncertainties of their actual positions may lead to drastic changes in the resulting dose distributions [118]. One major source of range and dose uncertainties in particle therapy resides in modeling the SPR of different tissues in the patient body. SPR are usually deduced using a calibration procedure based on CT images [118]. To reduce tissue conversion uncertainties, efforts are made to integrate dual energy CT (DECT) images aiming to resolve ambiguities in SPRs calibration from single energy CT (SECT) images, either by synthesizing DECT from SECT or by deriving SPR directly from DECT as discussed in the works of Wang et al. and Charyyev et al. [11,15]. Another approach to improve the accuracy of particle therapy dose calculations is applying AI to predict MC-equivalent dose distributions from analytical PB dose distributions, offering fast reduction of dose calculation errors [7]. Patient anatomical changes pose another major source of uncertainties, calling for adaptive particle therapy. AI can play a crucial role to monitor changes in patient anatomy and subsequently modify treatment plans when necessary [93]. On board CBCTs have been adopted to track the changes in patient anatomy. However, since CBCT images cannot directly be used for dose calculation, AI methods have been developed to solve this problem either by converting CBCT images to CT images [57,59,63,64] or by converting CBCT images directly to SPR maps [62]. Various AI based online verification methods intend to achieve in vivo assessment of the ion beam range and dose delivery have also been developed to deal with the uncertainties caused by patient anatomy changes, by deriving dose images from positron emission tomography (PET) [75-78], prompt gamma (PG) [95], secondary electron bremsstrahlung (SEB) X-ray images [102], acoustic signals [108], and luminescence images of water [109]. Besides the aforementioned work, AI techniques have also been applied to aid in MR-only treatment planning [23,24] and different parts of the patient treatment workflow including dose calculation [7,54,55], patient specific QA [112,113,116] and clinical decision systems [117]. Promising results reported in many of these reviewed studies demonstrate how AI-based methods can help exploit the intrinsic physics advantages of particle therapy. Among all the reviewed studies in this work, calculating the deposited dose by particles, as accurate as the gold standard MC simulations, can be interpreted as one of the most challenging areas in terms of applicability of AI in particle therapy. Unlike photons, the steep distal fall of particle dose distributions makes them very sensitive to anatomical changes. In addition, the effect of each heterogeneity along the trajectory of the proton does not manifest instantly but appears at the end of their range [55]. This would coerce a full 3D processing of the entire CT data [54] and the resulting GPU memory shortages, or a sequenced representation of clipped areas around single pencil beams [55]. Moreover, the challenges in performing dose calculation tasks can be even more extreme when is performed for estimating the deposited dose by heavy-ions such as Carbons. Here, the nuclear models describing the underlying secondary particle generations are uncertain, not to mention the difficulty of accurate estimation of the deposited "tail" dose resulted from these secondary particles. Future works related to AI applications in other important research areas of particle therapy like robust/biological/4D optimization, automated treatment planning and quality assurance are worth exploring.

Despite the promising results, AI based methods have not yet enjoyed widespread implementations in particle therapy clinical routine. One of the main criticisms is the lack of interpretability in AI models. The concern is that AI based models are “black boxes” in which the model extracts knowledge and information from input data and in which predictions are made without human understanding [4]. This issue is particularly important in AI applications in particle therapy as the resulting models may potentially influence the clinical decision and thereby the patient treatment outcome [119]. Another major challenge in the field of AI application in radiotherapy is that the accuracy of model outputs directly depends on the quality and quantity of input data. Indeed, it is usually the training data, rather than the algorithm selection or parameters settings, that most profoundly affect the model’s performance. This implies that AI models need to be trained with a sufficient variety of representative examples to be able to capture underlying structure of the data that allows the models to generalize to new cases [4]. However, data collection as well as data pre-processing is often time-consuming and usually requires certain domain knowledge from medical professionals [119]. Moreover, data from different institutions can represent different distributions in the latent space, which may cause a model trained on one institution's data to fail to generalize to a different one. Therefore, to generate better AI models that will become clinical standard, joint efforts should be made by researchers and clinicians from different institutions to standardize data acquisition and processing protocols and build coherent and sufficiently large databases in the field of particle therapy. AI applications in particle therapy must be clinically validated and its generalizability must be carefully checked before being introduced into routine clinical practice.

In general, the biggest challenges of AI application in particle therapy remain the lack of sufficient training data and acceptance of a black box that guides clinical decisions. However, the advantages that AI can offer in pattern detection will make it difficult to ignore. Likely first applications that enter the clinic will be in a helper function, providing comparisons to previously treated patient cohorts and plan suggestions or guidance, auto-detection of anatomical contours and secondary QA options. Once experience of interactions with AI become standard, the next level of integration will start, including corrections of patient images from CT, CBCT or MRI to predict SPR maps or dose distributions. Another driver to bring AI into clinical routine may be the joint integration of various methods and technologies discussed in this paper for the purpose of a seamless adaptive radiotherapy system where the physical advantages of particle therapy can be fully exploited via an accurate delivery of particles to the dynamic anatomical target.

To conclude, AI methods have been successfully applied to help exploit the physics advantages of charged particles. AI methods have great potential to improve the accuracy and efficiency of tasks in treatment workflow in particle therapy. However, a lot of work remains to be done to address the remaining challenges, most notably the lack of training data and acceptance of the "black box', before AI applications in particle therapy enjoy widespread implementation in clinical practice. Despite these challenges, AI applications are particularly useful for particle therapy due to the higher requirements in accuracy for the calculated dose distributions. We therefore expect AI to play an important role in the future of particle RT.

*Acknowledgment*

We would like to thank Dr. Jonathan Feinberg for editing the manuscript.